# A Survey of Blockchain Applications in Different Domains

Wubing Chen
Xi'an Jiaotong University
28 Xianning W Rd, ErHuan Lu
Xian Shi, China 710048
+86 13772529952
wuzbingchen@gmail.com

Zhiying Xu, Shuyu Shi
Nanjing University
22 Hankou Rd, Gulou Qu
Nanjing, China, 210008
+86 17763200154, +86 17761719569
maximilian.xu.2015@gmail.com,
ssy@nju.edu.cn

Yang Zhao, Jun Zhao
Nanyang Technological University
50 Nanyang Ave
Singapore 639798
+65 86483534
S180049@e.ntu.edu.sg,
JunZhao@ntu.edu.sg

## ABSTRACT
Blockchains have received much attention recently since they provide decentralized approaches to the creation and management of value. Many banks, Internet companies, car manufacturers, and even governments worldwide have incorporated or started considering blockchains to improve the security, scalability, and efficiency of their services. In this paper, we survey blockchain applications in different areas. These areas include cryptocurrency, healthcare, advertising, insurance, copyright protection, energy, and societal applications. Our work provides a timely summary for individuals and organizations interested in blockchains. We envision our study to motivate more blockchain applications.

## 1. INTRODUCTION
Blockchains are now talked about in the news worldwide. They have already been adopted in many applications of various domains as decentralized approaches to fraud-resistant computing without a trusted authority. A blockchain is a distributed, append-only log of time-stamped records that is cryptographically protected from tampering and revision [1].

In this paper, we survey blockchain usage in different areas including cryptocurrency, healthcare, advertising, copyright protection, energy, and societal applications. Some examples are discussed below, with more presented in the next section. Since the first blockchain-based cryptocurrency named bitcoin was introduced in 2008 [1], various other cryptocurrencies have emerged, including Litecoin, Namecoin, SwiftCoin, and Bytecoin [2]. With blockchains, fragmented healthcare records are combined to track personal health records [3]. MetaX [4] uses blockchains to counter fraud in digital advertising. InsureX [5] is a blockchain-based insurance marketplace to improve efficiency in insurance industry.

For individuals and organizations interested in blockchains, our paper provides a timely summary. Moreover, our work will motivate more blockchain applications. Different of our work discussing applications in different areas, prior survey papers on blockchains focus on either technical issues such as security [6] and consensus protocols [7], or specific applications such as the Internet of Things [8] and banking [9].

The rest of the paper is organized as follows. In Section 2, we elaborate blockchain applications in different areas. Section 3 presents related work. We conclude the paper in Section 4.

## 2. BLOCKCHAIN APPLICATIONS
### 2.1 Cryptocurrency
One of the most active areas of the blockchain is in the financial sector, especially in the field of cryptocurrency. Since the advent of the first carrier bitcoin in the blockchain [1], various cryptocurrencies have emerged. Because Bitcoin's anonymity, verifiability, decentralization and consensus mechanisms are characteristic, the value of Bitcoin has now reached an astonishing $6,300 per BTC [9]. At the same time, some other cryptocurrencies with more improved features have emerged, and constitute the current booming cryptocurrency market. Among them, Ethereum [10] created a public blockchain platform in which smart contracts can be deployed in 2015. With the emergence of contracts, the blockchain technology is applied to a wider range of business scenarios including contract processing, ownership changes, the Internet of Things, and the sharing economy [11]. In addition to being used in the field of cryptocurrencies, the blockchain is increasingly being used in financial services, including stock exchanges, cross-border payments, repurchase agreements, and digital identities. Utilizing the nature of the blockchain's distributed transaction ledger, the Bank of England Santander [12] used the technology provided by the payment protocol and exchange network based on Ripple to transfer payments in real time through a mobile application. The Australian Securities Exchange [13] claimed that Bitcoin technology would be used to replace the current clearing system with the aim of reducing transaction costs and making transactions faster and safer. Oxygen [14], a trade company based in London, announced the launch of its repurchase blockchain platform (Repos). When the repurchase agreement is initiated, the bank and the borrower respectively send their money and collateral to the pre-defined smart contract address, which locks in the circulation of the collateral, deposits the money into the

borrower's account, and continues to track all transactions.The trend toward electronic finance makes it natural for the blockchain and makes repurchase an automated clearing house. Many banks have also begun to invest in blockchains in recent years. Fidor Bank [15] is an online bank in Germany and the first mainstream bank to experiment with virtual currency and blockchains. In cooperation with a bitcoin exchange Karken headquartered in San Francisco, the exchange of euros and bitcoins was launched in October 2013. Fidor Bank works with Ripple Labs to provide low-rate transfer services using the payment technology of the other party. In February 2015, the bank partnered with bitcoin.de to launch the P2P Bitcoin Transfer Service. Citibank has built three independent systems based on blockchain distributed technology internally. The blockchain technology is also one of the five key areas of focus that Citigroup released in July 2015. Although the decentralized trust mechanism of blockchain technology can better solve the problem of value exchange on a global scale. However, from the current actual operation, there are still limitations in the following aspects. The first is the security of the blockchain. The blockchain is different from traditional financial facilities. Traditional financial facilities are controlled by an organization, and related software and hardware facilities are not publicly available. But the blockchain is an open application, and the code for the blockchain system is shared among the participants. As a result, blockchain-based applications are more vulnerable to attack than traditional financial facilities. The second is the privacy protection of the blockchain. In the traditional financial business model, data is stored on the central server, and the system operator protects data privacy. In blockchain-based applications, data is publicly transparent and each participant is able to get a complete data backup. Despite the existence of "pseudo-anonymous" in blockchains, for financial institutions, in some financial business scenarios that must be kept secret, this model is too simple to accommodate the needs of complex financial services.

## 2.2 Healthcare

Healthcare is also ready to invest in blockchains as new business cases emerge. Several characteristics provided by the blockchain are of great interest to healthcare organizations: disintermediation, transparency, auditability, industry collaboration, and new business models [16]. The scattered medical records caused by the transfer between different medical institutions have become a core problem that hinders healthcare IT. The blockchain provides an opportunity to create a platform for reliable recording. With the blockchain, highly fragmented healthcare records can be combined to provide an opportunity to track personal health records [3]. At the same time, access to healthcare records is an ethical dilemma. A major problem with this kind of application is to form the foundation of a high integrity tracking capability. Privacy concerns, fragmentation and inherent complexity of medical records make history-based diagnosis expensive. With blockchain, continuous tracking of the flow of services and money could be realized at a lower cost. Matthews [17] claims that the combination of artificial intelligence and blockchain could lead to solutions for healthcare problems. However, to achieve these wonderful visions, in addition to technical barriers such as accessing and storing data to the blockchain, there are also barriers from policy and privacy issues.

## 2.3 Advertising

As a distributed, immutable, transparent ledger, the blockchain is a natural fit for the digital advertising supply chain. Advertisements fraud, inefficiency, and lack of transparency have always been core issues that digital advertising needs to solve. With the help of blockchains, we can significantly improve efficiency and transparency, reduce costs, and prevent fraud [18]. Digital advertiser spent $209 billion [19] on digital in 2017 worldwide. Fueled by a large influx of venture capital during the past two years, many companies have been actively building blockchain-based advertising systems [20]. A blockchain company MetaX [4] aims to address issues of fraud and lack of transparency that impact digital advertising using blockchain technology. Some applications have been constructed for public testing. One of these applications is called Ads.txt Plus, which is designed to root out fraudulent sellers and resellers across the programmatic supply chain as an open source software. Premion [20], a division of TEGNA which is a leading over-the-top (OTT) advertising platform, has partnered with technology company MadHive to develop a transaction platform for OTT using blockchain. At the same time, MadHive has launched MAD Network with a vision for blockchain in OTT that is much grander. As a decentralized application, MAD builds trust between all parties in the advertising value chain as blockchain nodes [20]. One of the largest MVPDs (Multichannel Video Programming Distributors) will deploy blockchain platforms in premium digital channels and video supply chains, where marketers, programmers and operators can use smart contracts to plan, target, and report on ad buys across digital, broadcast, and streaming. In addition to improvements in quantifiable metrics, the blockchain can also improve the consumer's user experience. By the lights of the blockchain, marketers can now build customer profiles directly from customers and get all the information they are willing to share in a simple move. Peer-to-peer interaction with no intermediation, means that brands can interact directly with customers and make the most of their information. Customers are not only those who pay for real money or tokens with goods or services, but also those who pay with time and attention. All aspects mentioned help to improve the marketing plan of any company. All in all, a decentralized, distributed, and Turing-complete blockchain network will greatly facilitate the development of the digital advertising market.

## 2.4 Insurance

Traditional insurance policies are often processed on paper contracts, which means claims and payments are error-prone and often require human supervision. Compounding this is the inherent complexity of traditional insurance, which are consumers, brokers, insurers and reinsurers, as well as insurance's main Product — risk. As a kind of distributed ledger of the blockchain, it improves insurance industry efficiency from 4 respects: fraud elimination, claims automation, data analysis with the Internet of Things (IoT), and Reinsurance [21]. With all data including personal historical credit information, accident environmental information, historical policy information all flows in the blockchain network, the inherent scalability of blockchain and the assistance of IoT, the insurance industry that sells risk will usher in a huge revolution. Some forerunners have started to enter this process. In October 2016, the Blockchain Insurance Industry Initiative (B3I) was jointly launched by the top five insurance giants to study the application feasibility of blockchain in the insurance industry and develop blockchain-based proofs of concept for insurance [11]. InsureX [5], the world's first blockchain-based alternative insurance marketplace, aims to solve inefficiencies in today's insurance industry. An insurance protocol on a blockchain, called Aigang [22] which would enable community, companies, developers build insurance prediction markets and insurance Products themselves were proposed. They will build a self-insurance platform for any manufacturer or insurance company with a smart contract and a risk-based tokenization system.

## 2.5 Copyright protection

The development of the Internet have been accompanied by copyright issues regularly. From peer-to-peer file-sharing services, such as Napster [23] and Grokster [24] to photographs on the Web, copyrights have not always been respected [25]. From the perspective of a file holder, copyrights are often ignored or under some attacks. Therefore, unauthorized (also illegal) file-sharing and use of copyrighted content remains a significant problem. Now, blockchain technology has brought some light to this issue.

Blockchain is a decentralized, distributed digital ledger of records [26]. Considering that a file is duplicated thousands of times across the network, this network is designed to regularly update and reconcile all the copies so that all records are consistent. No single computer or organization is responsible for the blockchain. The property of no central storage location makes it almost impossible to manipulate or corrupt. Started with the base record, all changes added to the ledger can never be altered. Therefore, whenever a file copyrighted is used illegally, a digital ledger holding the owner's information and detailed transaction history is truly public and easily verifiable.

Take images online as an example. One of the biggest problems in copyrights protection is policing the use of images by who place them online. For photographers, platforms are springing up to provide mechanisms where their images can be uploaded to a blockchain and not only services of authentication for the ownership, but a means to police unauthorized use are provided. For example, Binded [27] touts itself as "the world's first copyright platform" for blockchain, creating "a unique fingerprint (cryptographic hash) for each copyright record". By tracking copyright records, Binded facilitates copyright protection by enabling access to the circulation path of copyright knowledge through blockchain. With a digital fingerprint at its fingertips, photograph owners can police online sites which use them, including media platforms lick Instagram and Twitter. Other service providers, such as COPYTRACK [28], are also blockchain-based copyright platforms.

However, there are also limitations on copyright protection with blockchain. One of the questions is initial authentication in the uploads. That is, how to prove ownership when a user uploads an image? Other limitations, such as issues inherent in monitoring the license usage of anonymous and untraceable buyers still need to be solved.

## 2.6 Energy

The trading of energy and commodities, even the simplest transactions, is often a balanced game of multi-parties [29]. From the execution to the conclusion of the transaction, both parties should coordinate and verify the transaction data. Additionally, through the transaction lifecycle, a company may need to interact with other counterparties, exchanges, brokers, logistics providers, banks, regulators and price reporters. In addition, the verification process needs to be carefully coordinated not only between the two parties of the transaction, but also within the company to maintain manual processes between different departments to ensure an accurate view of the entire transaction process.

Based on blockchain technology, it is possible to simplify both the internal workflows and the processes with external markets, which can completely change the arrangements of energy transactions. Also, streamlining the above processes will trigger notable savings (e.g., cutting down labor works, manual and semi-automatic costs, reducing capital costs by speeding up the settlements and technological efforts via abating dependence on a lot of systems).

Overall, the applications on energy trading we discovered fell into the following major categories:

**Energy trading markets:** Within this category, two camps have emerged: some initiatives aim to use blockchain to fundamentally rearrange the existing energy system while others seek to incrementally improve it. For example, Peer-to-peer cryptocurrency transactions seem to naturally give the blockchain a non-central character. In other areas, including the energy sector, many researchers and developers take for granted the idea of decentralization. We are pessimistic about this because they are trying to subvert only centralized management rather than a more efficient way of managing complex systems. Another example is grid transactions. These projects include restructuring existing power wholesale markets, through blockchain, where transactions will be verified faster and at less cost. While reforming the traditional electricity market, the new distributed energy market will also be likely to flourish and become a "power mining machine". Once the large-scale and complex power market changes like this, it will greatly promote the development of power industry and improve productivity.

**Energy Financing:** Many enterprise plans have intended to apply blockchain technology, including utilizing cryptocurrencies to raise capital, mostly concentrating on the green energy field. Blockchain will make it easier to raise capital for the clean energy projects via connecting more potential investors. Nonetheless, it is not obvious yet whether it is really indispensable to build up a decentralized network for accelerating the process of raising funds.

**Sustainability Attribution:** One of the most intuitive applications of blockchain in the energy industry is to continuously record energy production processes, including proportion of renewable energy, and pollution emission data. These honest records help improve fraud and poor decision-making.

**Electric Vehicles:** Even up to now, there are still fundamental obstacles before the popularization of electric vehicles (EVs), including the scarce charging infrastructures and the complexity of deploying them. Blockchain technology makes it convenience for personal owners of these infrastructures to provide services. Also, it can simplify the process, thus cutting down the costs during the charging. Obviously, these advantages could draw us nearer to the future of wide applications of EVs. Besides that, it is possible for EVs to charge and discharge by analyzing the needs of electricity and applying smart contracts, and serve as batteries to stabilize the allocation of energy.

For now, the main limitation within blockchain when applied on energy business lies on performance [30]. Blockchain applications such as Bitcoin [1] have limited transaction capacity, currently it is three transactions per second; seven per second are the maximum. Another limitation is critical mass. To use blockchain as a common industrial infrastructure, common standards need to be agreed within the industry. It's going to be a huge challenge because there are so many parties that need to agree.

## 2.7 Society Applications

**Non-traditional money lending:** Smart contracts, the next-generation network infrastructure expected to solve credit problems, can upend traditional borrowing relationships. In the traditional lending relationship, the lender not only lends money but also takes risks, which also leads to the high loan interest and the mortgage of goods in the traditional lending relationship, and the value of goods mortgage is often higher than the loan amount. With smart contracts, borrowers can use virtual assets as collateral, not only to prevent discounts on physical items, but also to reduce the credit cost. There is no need to show the lender credit or work history and numerous documents. The property is encoded on the blockchain for all to use.

**Car / smartphone:** For example, a car key with an anti-theft device can only be activated when you click on the correct protocol on the key. The smartphone will only work if you enter the correct password. They are all committed to encryption technology to protect ownership. The problem with the original form of intelligent property is that the key is kept in a physical container and cannot be easily transferred or copied. The blockchain ledger solves this problem by allowing blockchain miners to replace and copy lost protocols.

**Blockchain music:** In both the record era and the digital music era, copyright issues have plagued music publishers. This problem can be solved by creating a traceable music copyright database via blockchain and smart contract. Further, you can even send revenue to both the copyright owner and the musician in real time as consumer behavior occurs. For music enthusiasts, they can pay in digital currency.

**Blockchain government:** Democrats and Republicans questioned the security of the voting system in the 2016 U.S [31]. election. With the blockchain and smart contracts, each individual can see his vote and the overall statistical process. In addition, a significant proportion of the annual government budget is used to verify the flow of funds, and the use of blockchain technology can greatly simplify the process. Blockchain can be self-managed by providing a platform for companies, foundations, government agencies and individual citizens. Individuals can ensure their will is reflected through blockchain.

## 3. RELATED WORK

In contrast to our work discussing applications in different areas, prior survey papers on blockchains address either technical issues such as security [6] and consensus protocols [7], or specific applications such as the Internet of Things [8] and banking [9].

In addition to the applications mentioned in this paper, blockchains have also been used in education [32], real estate industry [33], transportation [34], charity [35], and food supply chain [36]. A future direction is to include these applications in a more detailed survey.

## 4. CONCLUSION

Blockchains have received much interest worldwide. In this work, we survey blockchain usage in different areas including cryptocurrency, healthcare, advertising, insurance, copyright protection, energy, and societal applications. Our paper presents a timely summary for entities with an interest in blockchains. Moreover, the discussion will motivate blockchain applications in more domains.

## 5. ACKNOWLEDGEMENTS

The research of Jun Zhao and Yang Zhao in this paper is supported by Nanyang Technological University (NTU) Startup Grant M4082311.020 and the project "Secure databases based on SGX" of Alibaba-NTU Singapore Joint Research Institute.

[15] Fidor Bank AG: The First Bank to Use the Ripple Protocol https://ripple.com/insights/fidor-bank-ag-the-first-bank-to-use-the-ripple-protocol/

[16] Blockchain for Healthcare https://www.ehidc.org/sites/default/files/resources/files/blockchain-for-healthcare-341.pdf

[17] Will Blockchain Transform Healthcare? https://www.forbes.com/sites/ciocentral/2018/08/05/will-blockchain-transform-healthcare/#2e63a56c553d

[18] Blockchain-based advertising services in smart cities. https://icorating.com/upload/whitepaper/tI26WlI7JTY7CVgjtTllhbMXABGqZHQk7YwxXVpx.pdf

[19] https://www.recode.net/2017/12/4/16733460/2017-digital-ad-spend-advertising-beat-tv

[20] Blockchain for Video Advertising: A Market Snapshot of Publisher and Buyer Use Cases https://www.iab.com/wp-content/uploads/2018/02/Blockchain_for_Video_Advertising_Publisher-Buyer_Use_Cases_2018-02.pdf

[21] The utility of Distributed Ledger Technology in facility arrangements for Brokers and (Re)Insurers https://static1.squarespace.com/static/5b8fab6996d455a32e992c8e/t/5b90e329aa4a9977bab7e87c/1536326606876/ChainThat+Facilities+white+paper

[22] Autonomous insurance network - fully automated insurance for IoT devices and a platform for insurance innovation built around data https://docs.google.com/document/d/1668B8uDH8JKCa7lp04AaSHI1XxiIcg6SLVb-doskX1g/edit?usp=sharing

[23] https://us.napster.com/

[24] http://www.grokster.com/

[25] https://abovethelaw.com/2018/02/how-blockchain-just-may-transform-online-copyright-protection/

[26] https://www.diyphotography.net/blockchain-copyright-protection-a-viable-solution-to-prove-ownership-of-creative-works/

[27] https://binded.com/

[28] http://www.copytrack.com/

[29] Merz, M. Potential of the blockchain technology in energy trading. Burgwinkel, Daniel. Blockchain Technology Introduction for Business and IT Managers. de Gruyter, 2016.

[30] Dütsch, G, and Steinecke, N. Use cases for blockchain technology in energy and commodity trading. Snapshot of current developments of blockchain in the energy and commodity sector. pwc, 2017.

[31] https://blockgeeks.com/guides/blockchain-applications/

[32] Sharples, M., and Domingue, J.. The blockchain and kudos: A distributed system for educational record, reputation and reward. In *European Conference on Technology Enhanced Learning*. 490-496, 2016.

[33] Spielman, A. *Blockchain: digitally rebuilding the real estate industry*. Doctoral dissertation, Massachusetts Institute of Technology (MIT), 2016.

[34] Yuan, Y., and Wang, F. Y. Towards blockchain-based intelligent transportation systems. In *IEEE 19th International Conference on Intelligent Transportation Systems (ITSC)*. 2663-2668, 2016.

[35] Jayasinghe, D., Cobourne, S., Markantonakis, K., Akram, R.N. and Mayes, K. Philanthropy on the Blockchain. In *IFIP International Conference on Information Security Theory and Practice*. 25-38, September 2017.

[36] Tian, F. An agri-food supply chain traceability system for China based on RFID & blockchain technology. In *IEEE 13th International Conference on Service Systems and Service Management (ICSSSM)*. 1-6, 2016.
21View publication stats